\documentclass[amsmath,amssymb,aps,pra,preprint,groupedaddress]{revtex4-1}

\usepackage{graphicx}
\usepackage{dcolumn}
\usepackage{bm}
\usepackage{color}

\begin{document}


\title{J-shaped stress-strain diagram of collagen fibers: Frame tension of triangulated surfaces with fixed boundaries}



\author{Yu Takano}
\affiliation{National Institute of Technology, Ibaraki College, 
Nakane 866, Hitachinaka, Ibaraki 312-8508, Japan}

\author{Hiroshi Koibuchi}
\email[]{koibuchih@gmail.com}
\affiliation{National Institute of Technology, Ibaraki College, 
Nakane 866, Hitachinaka, Ibaraki 312-8508, Japan}

\date{\today}

\begin{abstract}
We present Monte Carlo data of the stress-strain diagrams obtained
using two different triangulated surface models. The first is the
canonical surface model of Helfrich and Polyakov (HP), and the second
is a Finsler geometry (FG) model. The shape of the experimentally
observed stress-strain diagram is called J-shaped. Indeed, the diagram
has a plateau for the small strain region and becomes linear in the
relatively large strain region. Because of this highly non-linear
behavior, the J-shaped diagram is far beyond the scope of the ordinary
theory of elasticity. Therefore, the mechanism behind the J-shaped
diagram still remains to be clarified, although it is commonly
believed that the collagen degrees of freedom play an essential
role. We find that the FG modeling technique provides a coarse-grained
picture for the interaction between the collagen and the bulk material.
The role of the directional degrees of freedom of collagen molecules or fibers
  can be understood  in the context of FG modeling. We also
discuss the reason for why the J-shaped diagram cannot (can) be explained by the HP (FG) model. 
\end{abstract}


\maketitle

\section{Introduction}
The mechanical properties of macroscopic membranes, such as human
skin, have been extensively studied experimentally for a long time
\cite{Biomat-2013,MCLS-PMatSAc-2008,Greven-etal-JMorph-1995}.  One
interesting mechanical property is the stress-strain diagram. This
diagram is called "J-shaped" because of its plateau (linear behavior)
in the small (large) strain region \cite{FMZRAB-JSBiol-1997,TDRW-JMCB2013,TFCC-CLINICS2010,RKSRV-ASME2002,SBVN-ABME2000}. This
J-shaped curve is quite different from the curve expected from the theory of elasticity, and it is also different from the curve observed in rubber elasticity \cite{Flory-1959}. Moreover, from an engineering perspective, this non-linear behavior attracts a considerable amount of attention for biomaterial functional technology \cite{JMBBM-2013,GVRB-NL2011}.  
For these reasons, many efforts have been devoted to understanding the origin of such a specific and unusual response to external forces.  However, the mechanism still remains unclear, although it is widely accepted that the internal structure such as the collagen degrees of freedom \cite{Maier-Saupe-1958,de-Gennes-1979,Doi-Edwards-1986}, the notion of collagen network \cite{JMBBM-2016,MCPolymer-PRE2008,Polymer-PRE2003,GJug-JNCS2014}, and the notion of fibers \cite{Fiber-RMP2010,Fiber-PRE2012}  play essential roles in the J-shaped behavior.

In this paper, we use surface models for membranes, such as  the Helfrich and Polyakov (HP) model \cite{HELFRICH-1973,POLYAKOV-NPB1986,Bowick-PREP2001,WIESE-PTCP19-2000,NELSON-SMMS2004,GOMPPER-KROLL-SMMS2004} and a Finsler geometry (FG) model \cite{Koibuchi-Sekino-PhysicaA2014,OK-POL2017}, to calculate the stress-strain diagram. 
The purpose of this study is to clarify the J-shaped behavior from the
perspective of the theory of two-dimensional surfaces, which undergo
thermal fluctuations. In such two-dimensional surface models, the
stress $\tau$ can be obtained as the frame tension of the surface that
spans the fixed boundaries
\cite{WHEATER-JP1994,Cai-Lub-PNelson-JFrance1994,David-Leibler-JPF1991}. We
will show that the J-shaped curve of $\tau$ can be obtained in the
context of the HP model. However, a linear response of $\tau$ against
the strain is also detected in an intermediate region of the bending
rigidity $\kappa$. This linear behavior at the low strain region
contradicts with the existing experimental data
\cite{MCLS-PMatSAc-2008,Greven-etal-JMorph-1995,FMZRAB-JSBiol-1997}. In
contrast, the J-shaped curve of $\tau$ can be obtained independently
of $\kappa$ in the FG model. From this result, we confirm
that the stress-strain diagrams obtained using the FG model are consistent with the existing experimental data. 

The FG model is an extension of the HP model; hence, the Hamiltonian
is composed of the Gaussian bond potential $S_1$ and the bending
energy $S_2$ \cite{Koibuchi-Sekino-PhysicaA2014}. Moreover, the
Hamiltonian includes the sigma model Hamiltonian $S_0$ for variable
$\sigma$, which represents directional degrees of freedom of  collagen
or some internal molecular structures such as liquid crystals (LCs). This
variable $\sigma$ plays an important role in the J-shaped curve of the
stress-strain diagram, just like the polymeric degrees of freedom in
the aforementioned collagen network and fiber models.  We also note
that the variable $\sigma$ in this paper corresponds to the one used
in the FG model \cite{OK-POL2017} to represent the directional degrees of freedom of LC molecules in 3D liquid crystal elastomers \cite{Wamer-Terentjev,V-Domenici-2012,Terentjev-JPCM-1999,K-F-MacMolCP-1998}. We also note that the FG model in this paper is identical to the one introduced in Ref. \cite{KS-IJMPC2016}, where the surface tension and string tension of membranes are calculated on spherical and disk surfaces. In this paper, we use a cylindrical surface for calculating the diagram; thus, both the boundary conditions and the results in this paper differ from those reported in Ref. \cite{KS-IJMPC2016}. 

\section{Model}
\subsection{Frame tension of cylindrical surface}\label{cylinder}
\begin{figure}[ht]
\centering
\includegraphics[width=9.5cm]{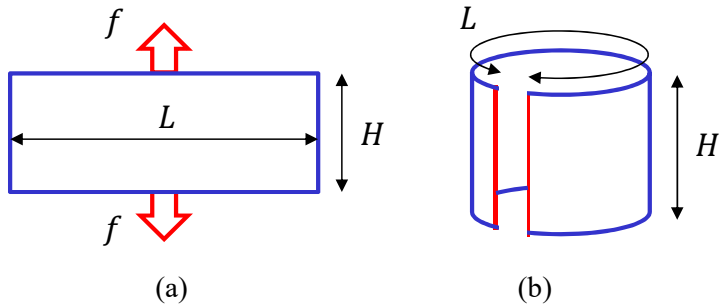}  
\caption{(Color Online) (a) An external force $f$ is applied to the square surface of the projected area $A_{\rm p}(=\!LH)$; (b) a cylindrical surface is made from the square surface by removing the vertical boundaries. } 
\label{fig-1}
\end{figure}
Let us assume that an external force $f$ is applied to a square
surface, the size of which is supposed to be $L\times H$
(Fig. \ref{fig-1}(a)). Let $\tau$ be the surface tension; then, we
have $f=\tau L$, and therefore, the accumulated surface tension energy
is given by $F=\int_{H_0}^{H} f dz=\tau L (H-H_0)=\tau A_{\rm p} +
{\rm const}$, where $A_{\rm p}=LH$ is the surface area. This surface
area $A_{\rm p}$ is the projected area of the frame, and therefore,
$A_{\rm p}$ is not always identical to the real surface area $A$ if
the surface is fluctuating. Thus, the surface tension $\tau$ is called
{\it frame tension} if  the real surface area deviates from the
projected area $A_{\rm p}$ due to the surface fluctuations
\cite{Cai-Lub-PNelson-JFrance1994,David-Leibler-JPF1991}. This frame
tension $\tau$ is the one that we would like to calculate in this
paper. In this paper, not only macroscopic membranes such as human
skin but also microscopic membranes are assumed as the research targets, where the thermal fluctuations are not always negligible.  
 
\begin{figure}[ht]
\centering
\includegraphics[width=9.5cm]{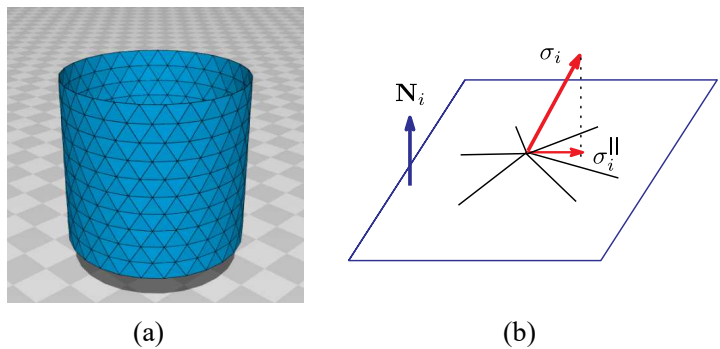}  
\caption{(Color Online) (a) A cylindrical surface of size $N\!=\!297$, which is made of the ruban of  $(N_1,N_2)\!=\!(11,27)$; 
(b) a unit normal vector ${\bf N}_i$ of the tangential plane at the vertex $i$, and the tangential component $\sigma_i^{\mid\mid}$ of $\sigma_i$. } 
\label{fig-2}
\end{figure}

A cylindrical surface is used for calculating the frame tension $\tau$
(Fig. \ref{fig-1}(b)). We use this cylindrical surface because the
cylinder has no boundary except the one to which an external force is
applied. A triangulated cylinder of size $N\!=\!297$ is shown in
Fig. \ref{fig-2}(a), where the height $H$ and the diameter $D$ are
assumed to be identical: $H\!=\!D$.  Let $N_1$ ($N_2$) be the total
number of vertices in the height (circumferential) direction; then, we
have $N\!=\!N_1\!\times\! N_2$, $H\!=\!\sqrt{3}(N_1\!-\!1) a/2$, and
$D\!=\!N_2a/\pi$, where $a$ is the triangle edge length. Thus, the
ratio $N_2/N_1\!=\!\sqrt{3}\pi/2$ is independent of the size $N$ (in
the limit of $N_1, N_2\to \infty$), and all cylinders that we use in the simulations are characterized by this ratio. 

\subsection{Finsler geometry model}
In this subsection, we introduce a FG model, which is identical to the
one introduced in Ref. \cite{KS-IJMPC2016}. The outlines of the
discrete model and the corresponding continuous model are shown in
this subsection and in Appendix \ref{FGmodel_Appendix}, respectively, in a self-contained manner.  First, we introduce the variable $\sigma_i (\in S^2: {\rm unit\;sphere})$ to represent the directional degrees of freedom of liquid crystal molecules (or collagen molecules). The Hamiltonian of the FG surface model is simply obtained by replacing the surface metric $g_{ab}$ with a Finsler metric (see Appendix \ref{FGmodel_Appendix}). To describe the interaction between the variables $\sigma$ themselves, we include the sigma model energy $\lambda S_0$ in the Hamiltonian $S$  with the interaction coefficient $\lambda$ such that
\begin{eqnarray}
\label{2D_discrete_S0}
&&S(\sigma,{\bf r})=\lambda S_0+S_1+\kappa S_2 + U_{B}, \nonumber \\
&& S_0(\sigma)= \left\{ \begin{array}{@{\,}ll}
                  -\sum_{ij}\sigma_i^{\mid\mid}\cdot \sigma_j^{\mid\mid} & \qquad ({\rm polar}) \\
                 -\left({3}/{2}\right)\sum_{ij}\left(\sigma_i^{\mid\mid}\cdot \sigma_j^{\mid\mid}\right)^2  & \qquad ({\rm nonpolar}) 
                  \end{array} 
                   \right., \\
&&
U_{B}= \sum_{i\in {\rm boundary}} U_{B}({\bf r}_i), \nonumber \\
&& U_{B}({\bf r}_i)=\left\{ \begin{array}{@{\,}ll}
                 \infty &  (|z_i-H| >\delta_B\; {\rm or}\; |z_i| >\delta_B) \nonumber\\
                  0 &  ({\rm otherwise}) 
                  \end{array} 
                   \right., \nonumber 
\end{eqnarray}
where $\sigma_i^{\mid\mid}$ (see Fig. \ref{fig-2}(b)) is defined as
\begin{eqnarray}
\label{sigma-parallel}
\sigma_i^{\mid\mid}=\sigma_i-(\sigma_i\cdot {\bf N}_i){\bf N}_i.
\end{eqnarray}
In $S_0$ of Eq. (\ref{2D_discrete_S0}), we assume the factor $3/2$ for the non-polar interaction, because Lebwohl-Lasher potential for LCs includes this factor,  although LCs are not always included in collagen fibers \cite{Leb-Lash-PRA1972}. In our FG model, the variable $\sigma$ represents the direction of collagen molecule as mentioned above. The collagen fiber is made of collagen fibrils, which is made of collagen molecules (a hierarchical structure), and therefore the fiber becomes relatively stiff \cite{GVRB-NL2011}. In addition, the fibers are loosely connected by cross-linkers. For these reasons, the collagen fiber networks are always locally ordered. This is in sharp contrast to the case of polymers, which have not only crystalline but also randomly disordered structure. Therefore, the collagen fiber networks change from locally ordered to globally ordered states when they are expanded by external tensile forces.  This coarse-grained picture of locally ordered structure of fiber network is expressed by the energy term $\lambda S_0$ with finite $\lambda$ for $\sigma$ in our FG model. The reason why the polar interaction is also assumed for $\sigma$ is simply to compare the results with those of non-polar interaction. 
 The coefficient $\lambda$ of $S_0$ is fixed to $\lambda\!=\!1$ in both polar and non-polar interactions in the simulations. The fact that $\lambda$ is fixed to  $\lambda\!=\!1$ is the cause of locally ordered configuration of $\sigma$, although $\lambda\!=\!1$  corresponds to the isotropic phase at least for small strain region.

The stiffness of the fibers can be measured by $EI$, where $E$ is the Young modulus and $I$ the second moment of area. This $EI$ is called bending rigidity, which measures stiffness of macroscopic elastic materials. In contrast, the bending rigidity $\kappa$ in Eq. (\ref{2D_discrete_S0}) corresponds to stiffness of microscopic membranes such as red cells. Thus, $\kappa$ in Eq. (\ref{2D_discrete_S0}) should be simply considered as a microscopic parameter that can be controlled depending on the rigidity of fibers in consideration.  

The vector ${\bf N}_i$ is the unit normal vector of the surface at vertex $i$, and it is defined as
\begin{eqnarray}
\label{normal_vect_on_vertex}
{\bf N}_i=\frac{\sum_{j(i)} A_{j(i)}{\bf n}_{j(i)}}{\left|\sum_{j(i)} A_{j(i)}{\bf n}_{j(i)}\right|},
\end{eqnarray}
where $A_{j(i)}$ and ${\bf n}_{j(i)}$ denote the area and the unit normal vector of the triangle ${j(i)}$ sharing the vertex $i$, respectively.  Note that $S_0$ is implicitly dependent on ${\bf r}$ because $\sigma_i^{\mid\mid}$ depends on the surface shape.  The expressions for the Gaussian bond potential $S_1$ and the bending  energy $S_2$ are
\begin{eqnarray} 
\label{2D_discrete_S1_S2}
&&S_1=\frac{1}{6}\sum_{\it \Delta} \left[ \gamma_{12} \ell_{12}^2 + \gamma_{23} \ell_{23}^2  + \gamma_{31} \ell_{31}^2\right], \nonumber \\
&&S_2=\frac{1}{6}\sum_{\it \Delta} \left[ \kappa_{12} \left(1-{\bf n}_0\cdot {\bf n}_1\right) + \kappa_{23} \left(1-{\bf n}_0\cdot {\bf n}_3\right) \right. \nonumber \\
&& \qquad\qquad\qquad + \left. \kappa_{31} \left(1-{\bf n}_0\cdot {\bf n}_2\right)\right],  \\
&&\gamma_{12}=\frac{v_{12}}{v_{13}} + \frac{v_{21}}{v_{23}}, \; \gamma_{23}=\frac{v_{23}}{v_{21}} + \frac{v_{32}}{v_{31}},  \; \gamma_{31}=\frac{v_{31}}{v_{32}}+ \frac{v_{13}}{v_{12}}, \nonumber \\
&&\kappa_{12}=\frac{v_{13}}{v_{12}} + \frac{v_{23}}{v_{ 21}},  \; \kappa_{23}=\frac{v_{21}}{v_{23}} + \frac{v_{31}}{v_{32}}, \; \kappa_{31}= \frac{v_{32}}{v_{31}} + \frac{v_{12}}{v_{13}}. \nonumber
\end{eqnarray}
The derivation of these expressions from the continuous Hamiltonians is shown in Appendix \ref{FGmodel_Appendix}. The symbol $\ell_{ij}$ is the length of bond $ij$, and $v_{ij}$ is given in Eq. (\ref{tangent_comp}) (see also Fig. \ref{fig-A1} in Appendix \ref{FGmodel_Appendix}).

Here we should comment on the boundary condition for the cylindrical surface. Because of the definition of $v_{13}$ in Eq.(\ref{tangent_comp}), the variable $\sigma_1$ at vertex $1$ on the boundary cannot be vertical to bond $13$ on the same boundary. In fact, if $\sigma_1\cdot {\bf t}_{13}\!=\!0$  then we have $v_{13}\!=\!0$, and  therefore $\gamma_{12}\to\infty$. This divergence of $\gamma_{12}$ implies that $\sigma_1$ never be vertical to the boundary. Therefore, to remove such unphysical repulsive interaction with respect to the direction of tensile forces, we assume that the boundary vertices are able to move into the horizontal direction only slightly within small range $\delta_B$. This constraint for the boundary vertices is defined by the potential $U_B$. The small value $\delta_B$ is given by the mean bond length, and therefore we have 
\begin{eqnarray} 
\label{small-height}
\frac{\delta_B }{H}\left(=\frac{\rm mean\; bond\; length}{\rm height\;of\;cylinder}\right) \to 0 \quad (N\to\infty). 
\end{eqnarray} 
This implies that the constraint potential $U_B$ is negligible in the limit of $N\!\to\!\infty$:
\begin{eqnarray} 
\label{UB-negligible}
U_B \to 0 \quad (N\to\infty), 
\end{eqnarray} 
while $\sigma$ can be vertical to the boundaries or parallel to the direction of tensile forces.

The discrete partition function $Z$ is given by
\begin{eqnarray} 
\label{Part-Func}
&& Z(\lambda,\kappa; L) =\nonumber \\
&&  \sum_{\sigma}\int \prod _{i=1}^{2N_2} d {\bf r}_i\prod _{i=1}^{N-2N_2} d {\bf r}_i \exp\left[-S(\sigma,{\bf r})\right],
\end{eqnarray} 
where $Z(\lambda,\kappa; L)$ denotes that $Z$ depends on the
parameters  $\lambda, \kappa$ and the height $L$ of the cylindrical
surface. $\int\prod _{i=1}^{2N_2} d {\bf r}_i$ denotes the multiple
$4N_2$-dimensional integration for the $2N_2$ vertices on the upper
and lower boundaries of the cylinder. The vertices on the boundaries
are prohibited from moving in the height direction, and hence, the
corresponding integration $\int d {\bf r}_i$ becomes a 2-dimensional
integration. In contrast, the vertices on the surface, except for those on the boundaries, are not constrained by the boundaries; therefore, $\int\prod _{i=1}^{N-2N_2} d {\bf r}_i$ is understood to be the $3(N\!-\!2N_2)$-dimensional integrations for those $N\!-\!2N_2$ vertices. 

\subsection{Formula for calculating stress-strain diagram }\label{stress_strain}
The surface position ${\bf r}$ is the variable that is integrated out in the partition function $Z$, and for this reason, $Z$ becomes invariant under the change of the integration variable such that ${\bf r}\!\to\! \alpha {\bf r}\;(\alpha\in {\bf R})$. This property is called the scale invariance of $Z$, and it is used for calculating the stress-strain curve \cite{WHEATER-JP1994}. 

The scale invariance implies that $Z$ should be independent of the scale parameter $\alpha$ \cite{Koibuchi-etal-JOMC2016};
\begin{eqnarray} 
\label{scale_indep_Z}
  \left. {d Z}/{d \alpha}\right|_{\alpha=1}=0.
\end{eqnarray} 
The scaled partition function is given by
\begin{eqnarray} 
\label{scaled_Z}
 && Z(\alpha; \alpha^{-2}A_{\rm p})=\nonumber \\
&&\alpha^{3N-2N_2}\sum_{\sigma}\int \prod _{i=1}^{2N_2} d {\bf r}_i\prod _{i=1}^{N-2N_2} d {\bf r}_i \exp\left[-S(\sigma,\alpha{\bf r})\right], \nonumber \\
  &&S(\sigma,\alpha{\bf r})=\lambda S_0+\alpha^2\gamma S_1 + \kappa S_2,
\end{eqnarray} 
where $\alpha^{-2}A_{\rm p}$ in $Z(\alpha; \alpha^{-2}A_{\rm p})$
denotes the dependence of $Z$ on $\alpha$ arising from the fact that
the projected area $A_{\rm p}$ is fixed. This dependence of $Z$ on
$\alpha$ implies that $Z$ can be considered to be a two-component function. Thus,  
from Eq. (\ref{scale_indep_Z}), we obtain $d Z/d \alpha=\partial Z/\partial \alpha \!+\! [\partial Z/\partial (\alpha^{-2}A_{\rm p})][\partial (\alpha^{-2}A_{\rm p})/\partial \alpha]=0$. Dividing Eq. (\ref{scale_indep_Z}) by $Z$ and using $\partial (\alpha^{-2}A_{\rm p})/\partial \alpha=-2A_{\rm p}\alpha^{-3}$, we have 
\begin{eqnarray} 
\label{scale_indep_Z_1}
3N-2N_2 -2\gamma \langle S_1\rangle - 2\frac{A_{\rm p}}{Z}\frac{\partial Z}{\partial A_{\rm p}}=0.
\end{eqnarray} 
The mean value of the Gaussian energy $\langle S_1\rangle$ on the left-hand side is obtained by Monte Carlo (MC) simulations. However, the problem here is how to evaluate the final term $(1/Z)\partial Z/\partial A_{\rm p}$. To calculate this term, we assume that the free energy $F$ is given by 
\begin{eqnarray} 
\label{free_enrgy}
F=\tau \int_{A_0}^{A_{\rm p}} dA=\tau(A_{\rm p}-A_0), 
\end{eqnarray} 
where $\tau$ is the frame tension as mentioned above. Because the corresponding partition function is given by $Z=\exp (-F)$, we finally obtain $\tau \!=\! \left({2\gamma \langle S_1\rangle - 3N+2N_2}\right)/({2A_{\rm p}})$. 

The problem now is how to obtain the projected area $A_{\rm p}$ in this $\tau$. Let $D$ and $H$ be the diameter and the height of the cylinder, respectively. Then, it is natural to define $A_{\rm p}$ as $A_{\rm p}=\pi D H$ if $D$ is uniform in the sense that $D$ is independent of the height position $h$ of the cylinder. Note that $D$ corresponds to $L/\pi$ for the cylinder, such as the one in Fig. \ref{fig-1}(b).
However, the diameter $D$ is expected to generally depend on $h$
because the cylinder is not a three-dimensional one but rather a two-dimensional surface, and therefore, $D$ at the height position $h\!=\!H/2$   may be different from $D$ at $h\!\simeq\!H$ or $h\!\simeq\!0$, for example. For this reason, we use the diameter $D_0(=\!H_0)$ of the initial surface, which is used for the simulations of $\tau\!=\!0$. Thus, the formula for $\tau$ is given by 
\begin{eqnarray} 
\label{frame_tension}
&&\tau=  \left(2\gamma \langle S_1\rangle - 3N+2N_2\right)/\left(2A_{\rm p}\right), \nonumber \\
&& A_{\rm p}=\pi D_0 H, \quad (\gamma=1).
\end{eqnarray}

\subsection{Monte Carlo technique }
The multiple-dimensional integrations in $Z$ are simulated using the
standard Metropolis Monte Carlo technique
\cite{Mepropolis-JCP-1953,Landau-PRB1976}. The update of the vertex
position ${\bf r}$ is performed with the probability ${\rm Min}[1,
  \exp(-\delta S)]$, where $\delta S\!=\!S({\rm new})\!-\!S({\rm
  old})$ with the new position ${\bf r}^\prime\!=\!{\bf r}\!+\!\delta
{\bf r}$. The small change $\delta {\bf r}$ is randomly distributed in
a small sphere (or circle) of radius $R_\delta$, which is fixed to
keep an approximately $50\%$ acceptance rate of ${\bf r}^\prime$. The
vertices on the boundaries are allowed to move only in the horizontal
plane (${\bf R}^2$), whereas the other vertices move in the
three-dimensional space ${\bf R}^3$. The variable $\sigma$ is updated
such that the new variable $\sigma^\prime (\in S^2)$ is completely
independent of the old $\sigma$. One Monte Carlo sweep (MCS) consists
of $N$ consecutive updates for ${\bf r}$ and $N$ consecutive updates
for $\sigma$. After a sufficiently large number of MCSs, measurements of
 physical quantities are performed every $1000$ MCSs.  All the simulations
 in this paper are performed on lattices of size $N\!=\!10584$.

The initial height  $H_0(=\!D_0)$ for the frame tension $\tau$ in
Eq. (\ref{frame_tension}) is determined such that the equilibrium
configurations satisfy $\tau\!=\!0$. From this definition of $D_0$ and
that of $A_{\rm p}$,  the frame tension $\tau$ in
Eq. (\ref{frame_tension}) is considered to be the nominal stress in
the sense that $\tau$ is independent of $\pi\langle D\rangle$ the real
length of the circumference of the cylinder. To be more precise,
$\pi\langle D\rangle$ can be identified using the real length of the
circumference of the cylinder only when the cylinder is sufficiently
smooth and  has no surface fluctuations.  In the case of 3D LCE, the
nominal stress is calculated with a constant sectional area, which is
independent of the height of the cylinder \cite{OK-POL2017}. In contrast, the projected area $A_{\rm p}$ for $\tau$ in Eq. (\ref{frame_tension}) is proportional to the height $H$ of the cylinder.  Note that the diameter $D_0 (=\!H_0)$ is not always identical to $\langle D\rangle$ (the symbol $\langle \;\rangle$ is not used henceforth, for simplicity); however, the deviation between $D_0$ and $D$ is expected to be small because the cylinder is constructed such that the diameter equals the height for $\tau\!=\!0$.

\section{Results }

\subsection{Snapshots}
\begin{figure}[ht]
\centering
\includegraphics[width=9.5cm]{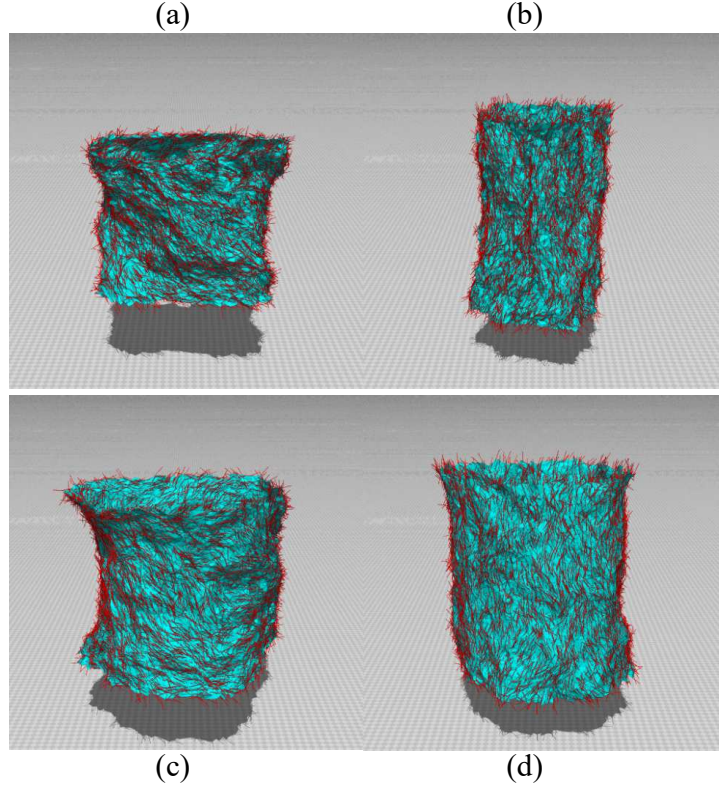}  
\caption{(Color Online) Snapshots of surfaces of non-polar model for  $\kappa\!=\!1$ with (a) $H/H_0\!=\!1$ and (b) $H/H_0\!=\!1.4$, and for  $\kappa\!=\!3$ with (c) $H/H_0\!=\!1$ and (d) $H/H_0\!=\!1.24$. The short lines (or burs) on the surfaces represent the variable $\sigma$. $N\!=\!10584$ and  $\lambda\!=\!1$. } 
\label{fig-3}
\end{figure}
First, we show snapshots of surfaces of non-polar model in Fig. \ref{fig-3}. From the snapshots, we confirm that the variable $\sigma$ is locally ordered when $H/H_0\!=\!1$ for $\kappa\!=\!1$ while it is globally ordered along vertical direction when $H/H_0\!=\!1.4$ (Figs. \ref{fig-3}(a),(b)). The reason why $\sigma$ is locally ordered is because the coefficient $\lambda$ of $S_0$ is fixed to $\lambda\!=\!1$.  
For the case  $\kappa\!=\!3$, we can also see almost the same ordering of $\sigma$ on the surfaces (Figs. \ref{fig-3}(c),(d)).

\subsection{Canonical model}
In this subsection, we present the results of the HP (or canonical) model and
discuss why the canonical model is insufficient for explaining the
existing experimental data of the J-shaped stress and strain diagram \cite{Biomat-2013,MCLS-PMatSAc-2008,Greven-etal-JMorph-1995,FMZRAB-JSBiol-1997}. The canonical model is defined as
\begin{eqnarray} 
\label{canonical_model}
 &&Z(\kappa; L) = \int \prod _{i=1}^{2N_2} d {\bf r}_i\prod _{i=1}^{N-2N_2} d {\bf r}_i \exp\left[-S({\bf r})\right], \nonumber \\
&&S({\bf r})=S_{-1}+\kappa S_3, \qquad\qquad\qquad ({\rm canonical})  \\
&&S_{-1}=\sum_{ij}\ell_{ij}^2, \quad S_3=\sum_{ij}\left(1-{\bf n}_i\cdot{\bf n}_j\right), \nonumber
\end{eqnarray} 
where $\ell_{ij}^2=\left({\bf r}_i-{\bf r}_j\right)^2$ is the bond
length squares
\cite{Bowick-PREP2001,GOMPPER-KROLL-SMMS2004,WHEATER-JP1994}. In
Eq. (\ref{canonical_model}),  we use the symbols  $S_{-1}$ and $S_3$
for the canonical Gaussian energy and the bending energy, respectively, to distinguish them from $S_1$ and $S_2$ in Eq. (\ref{2D_discrete_S0}) for the FG model. Note that these $S_{-1}$ and $S_3$ are not assumed as Hamiltonians in the FG model; however, these quantities can be obtained (or calculated) from the surface configurations of the FG model.  

\begin{figure}[ht]
\centering
\includegraphics[width=9.5cm]{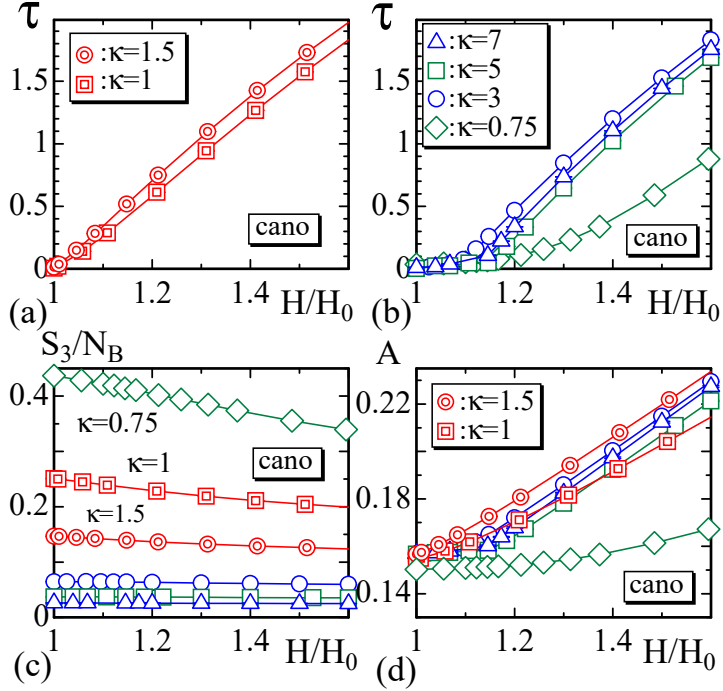}  
\caption{(Color Online) The stress $\tau$ vs. strain $H/H_0$ of the canonical model for (a) $\kappa\!=\!1$, $\kappa\!=\!1.5$ and  (b) $3\!\leq\!\kappa\!\leq\!7$,  $\kappa\!=\!0.75$;   (c) the bending energy $S_3/N_B$ vs. $H/H_0$;  and (d) the mean triangle area $A$ vs. $H/H_0$, where $S_3$ is defined in Eq. (\ref{canonical_model}). $N\!=\!10584$.  } 
\label{fig-4}
\end{figure}
As shown in Fig. \ref{fig-4}(a), $\tau$ is linear with respect to
$H/H_0$ for $\kappa\!=\!1.5$ and  $\kappa\!=\!1$. This figure also
shows that the shape of $\tau$ changes from linear to J-shaped when $\kappa$ increases from $\kappa\!=\!1.5$ to $\kappa\!=\!3$,  $\kappa\!=\!5$ and $\kappa\!=\!7$. It is also observed that $\tau$ becomes J-shaped when $\kappa$ decreases from  $\kappa\!=\!1.5$ to $\kappa\!=\!0.75$. To evaluate the surface smoothness, we plot the bending energy $S_3/N_B$  in Fig. \ref{fig-4}(c), where $N_B$ is the total number of bonds excluding the bonds on the boundaries on which $S_3$ is not defined. We find that the plateau of $\tau$ can be observed on relatively smooth surfaces of $S_3/N_B\!\leq\!0.07$ and on relatively wrinkled surface of $S_3/N_B\!\geq\!0.4$. On the surfaces of $S_3/N_B\!\simeq\!0.2$,  $\tau$ becomes linear with respect to $H/H_0$. The mean triangle area $A$, which is defined as
\begin{eqnarray}
 A=(1/N_T)\sum_{\it \Delta}A_{\it \Delta},
\end{eqnarray} 
is also J-shaped  if the corresponding $\tau$ is J-shaped
(Fig. \ref{fig-4}(d)), where $N_T$ is the total number of
triangles. Only for $\kappa\!=\!1.5$ is the mean triangle area $A$ almost linear. The J-shaped behavior of $A$ implies that the real surface area remains unchanged  for the small $H/H_0(>\!1)$ region, whereas the projected area $A_{\rm p}$ always changes linearly with respect to $H/H_0$.
 
Thus, we observe that the behavior of $\tau$ for
$\kappa\!\simeq\!0.75$ and $3\!\leq\!\kappa\!\leq\!7$ appears
J-shaped, and hence, this observation indicates that the J-shaped
diagram can be understood within the context of the canonical
model. However, the problem is the linear behavior of $\tau$ observed
in the intermediate region  $\kappa\!\simeq\!1.5$.  These results also
imply that $\tau$ has J-shaped behavior at low and high temperatures,
whereas it has a linear behavior at intermediate temperatures because
$\kappa$ has units of $k_BT$. To summarize these results, the linear $\tau$ at $\kappa\!\simeq\!1.5$ conflicts with the existing experimental results \cite{Biomat-2013,MCLS-PMatSAc-2008,Greven-etal-JMorph-1995,FMZRAB-JSBiol-1997}, at least in the context of the canonical model. 

The problem is why linear behavior is observed only at intermediate region of $\kappa$. 
 We first note that the linear behavior of $\tau$ for the large $H/H_0$ region is easy to understand. For the large $H/H_0$ region, the surface area is increased to a sufficiently large value,  while the bending energy $S_3$ is negligible compared with $S_{-1}$, which has  units of length squares;  hence, the energy supplied by the external force is accumulated only in $S_{-1}$. The interesting region of $H/H_0$ is close to $H/H_0\!=\!1$, where the surface can fluctuate if $\kappa$ is not very large. 
For the small $\kappa$ region, the surface is sufficiently wrinkled,
and therefore, the surface height $H$ is increased without changing
the bond length for $H/H_0$ close to $H/H_0\!=\!1$. 
 On such rough surfaces,  the long wavelength  mode of surface fluctuations is not expected. This means that the persistence length $\xi$ is relatively short, and hence, the external force at one of the two boundaries has no influence on the other boundary.  However, when $\kappa$ is increased to $\kappa\!\simeq\!1.5$, the surface becomes relatively smooth such that the long wavelength modes (or long-range correlations of surface fluctuations, such as surface normals) are expected to appear on the surface, and therefore, the external force applied on the boundary influences the entire surface such that $S_{-1}$ can be increased even for the small $H/H_0$ region. This is a reason for why no plateau is observed in $\tau$ in the intermediate region of $\kappa$. For the large $\kappa$ region, such as $\kappa\!\simeq\!5$, the surface is further smoothed, and the surface fluctuation is suppressed. On these relatively smooth surfaces with small surface fluctuations, the boundary effect is not mediated as the surface fluctuation modes for the small $H/H_0$ region, and this makes a plateau in $\tau$.  

Thus, the linear behavior of $\tau$ for the intermediate region of
$\kappa$ is typical of the two-dimensional fluctuating surfaces;
however, this linear behavior is unsatisfactory from the perspective
of the experimental fact that $\tau$ is J-shaped in biological
membranes
\cite{Biomat-2013,MCLS-PMatSAc-2008,Greven-etal-JMorph-1995,FMZRAB-JSBiol-1997}. 
The main reason for this is that the surface fluctuations are expected
 in the canonical surface model while these are suppressed in the macroscopic membranes such as skins and collagen fiber networks. 
We have to conclude that the canonical surface model is insufficient for explaining the J-shaped stress-strain diagram of macroscopic membranes. 
\subsection{FG model for $\lambda=0$}
\begin{figure}[ht]
\centering
\includegraphics[width=9.5cm]{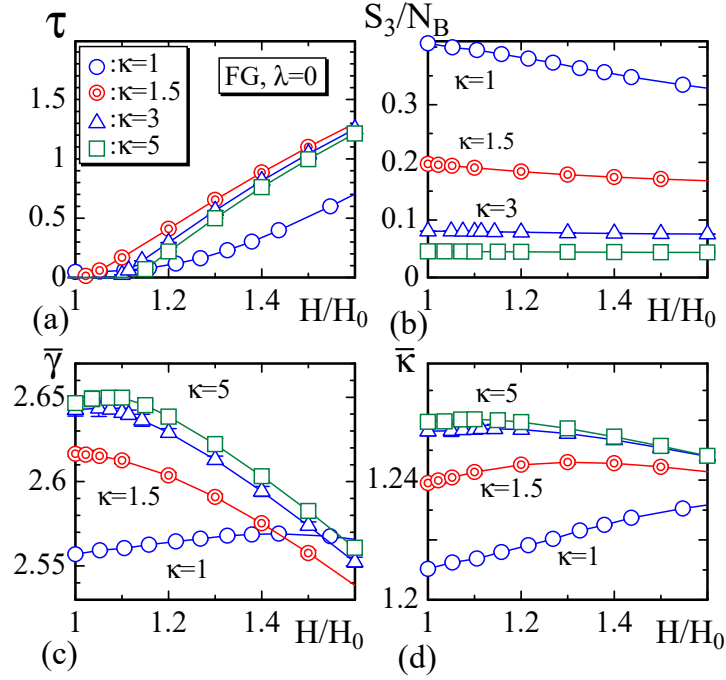}  
\caption{(Color Online) MC data of the FG model for $\lambda\!=\!0$:
  (a) $\tau$ vs. $H/H_0$, (b) $S_3/N_B$ vs. $H/H_0$, (c) $\bar \gamma$  vs. $H/H_0$, and (d) $\bar \kappa$  vs. $H/H_0$. $N\!=\!10584$. } 
\label{fig-5}
\end{figure}
In this subsection, we evaluate the equivalence between the canonical model and the FG model for $\lambda\!=\!0$. When $\lambda$ is zero in the FG model, the variable $\sigma$ becomes random, and hence, no anisotropy is expected on the surface \cite{Koibuchi-Sekino-PhysicaA2014,OK-POL2017}. Indeed, the FG model is an extension of the canonical HP model in the sense that the outputs of the FG model for $\lambda\!=\!0$ are consistent with those of the canonical model. 

Figure \ref{fig-5}(a) shows $\tau$ vs. $H/H_0$ of the FG model for
$\lambda\!=\!0$ with several different values of $\kappa$. As shown in
this figure, $\tau$ changes linearly against $H/H_0$ for $\kappa\!=\!1.5$, and $\tau$ is J-shaped for $\kappa\!=\!5$,  $\kappa\!=\!3$ and $\kappa\!=\!1$. From these results, it is clear that the dependence of $\tau$ on $H/H_0$ of the FG model for  $\lambda\!=\!0$ is consistent with that of the canonical model. Indeed, we find from $S_3/N_B$ shown in Fig. \ref{fig-5}(b) that the stress $\tau$ for the surfaces of $S_3/N_B \!\simeq\!0.2$ ($S_3/N_B \!\leq\!0.07$ and $S_3/N_B \!\geq\!0.35$) behaves linearly (has a plateau) with respect to $H/H_0$. This result is almost consistent with the results of the canonical model shown in Figs. \ref{fig-4}(a),(b),(c). 

Note that the role of  $\kappa$ in the canonical model is not always
the same as that in the FG model. Indeed,  $\kappa\kappa_{ij}$ plays
the role of the bending rigidity in the FG model, whereas  the
constant $\kappa$ is the bending rigidity in the canonical
model. Moreover, the surface tension coefficient $\gamma$ is fixed to
$\gamma\!=\!1$ in the canonical model, whereas in the FG model,
$\gamma_{ij}$ plays the role of the surface tension coefficient. For these reasons, to clarify the relation between the coefficients in the canonical model and those in the FG model, we calculate the mean values of $\gamma_{ij}$ and  $\kappa_{ij}$ such that
\begin{eqnarray}
\label{mean_k_g}
&&\bar \gamma=\frac{\sum_{ij}\gamma_{ij}\ell_{ij}^2}{ \sum_{ij}\ell_{ij}^2}=\frac{3S_1}{S_{-1}}, \nonumber \\
&&\bar \kappa=\frac{\sum_{ij}\kappa_{ij}\left(1-{\bf n}_i\cdot{\bf n}_j\right)}{\sum_{ij}\left(1-{\bf n}_i\cdot{\bf n}_j\right)}=\frac{3S_2}{S_3}. 
\end{eqnarray}
The reason for multiplying  $S_1$ and $S_2$ by a factor of $3$ in Eq. (\ref{mean_k_g}) is as follows. The sum of triangles $\sum_{\it \Delta}$ in $S_1$ and $S_2$ in Eq. (\ref{2D_discrete_S1_S2}) can be replaced by the sum of bonds $2\sum_{ij}$ because every term $\ell_{ij}^2$ or $1-{\bf n}_i\cdot{\bf n}_j$ is summed over twice in the sum $\sum_{\it \Delta}$. From this factor $2$, the factor $1/6$ in $S_1$, and the expression of $S_2$ in Eq. (\ref{2D_discrete_S1_S2}), we include the factor $3$ in $\bar \gamma$ and $\bar \kappa$ in Eq. (\ref{mean_k_g}). 

We observe that the dependence of $\bar \gamma$ on $H/H_0$ in
Fig. \ref{fig-5}(c) appears similar to that of $\tau$ in Fig. \ref{fig-5}(a). Indeed, both $\bar \gamma$ and $\tau$ have a plateau (linear behavior) for $\kappa\!=\!5$, $\kappa\!=\!3$ and $\kappa\!=\!1$ ($\kappa\!=\!1.5$). We also observe that the value of $\bar \gamma$ is in the range $2\!\leq\!\bar \gamma\!\leq\!2.8$, and it is larger than that of $\gamma(=\!1)$ of the canonical model. However, $\bar \gamma$ is included in $S_1(=\!\bar \gamma S_{-1})$, and therefore, we expect that this difference between $\bar \gamma$ and $\gamma(=\!1)$ does not make any difference between  the stresses $\tau$ of the canonical and FG models. Indeed, $\tau$ in Fig. \ref{fig-5}(a) for each $\kappa$ is almost comparable to (or slightly smaller than) $\tau$ in  Figs. \ref{fig-4}(a),(b) of the canonical model. 

Using the $\bar \gamma$, $\bar \kappa$ in Eq. (\ref{mean_k_g}) and the
Hamiltonians $S_{-1}$, $S_3$ in Eq. (\ref{canonical_model}), we have
the effective Hamiltonian for the FG model such that $S_{\rm eff}\!=\!\bar \gamma S_{-1}\!+\!\kappa\bar \kappa S_3$, which can also be written as $S_{\rm eff}\!=\!\bar \gamma\left[S_{-1}\!+\! \kappa(\bar\kappa/\bar \gamma) S_3\right]$.  In these expressions of $S_{\rm eff}$, the multiplicative factor 3 is dropped for simplicity. Furthermore, because the factor $\bar \gamma$ in $S_{\rm eff}$ can be dropped due to the scale invariance of $Z$, we finally have $\tilde S_{\rm eff}\!=\!S_{-1}\!+\! \kappa(\bar\kappa/\bar \gamma) S_3$. 

From this $\tilde S_{\rm eff}$, the differences between $S_3/N_B$ of
the canonical and FG models are understood. As shown in
Fig. \ref{fig-5}(d), $\bar \kappa$ is slightly larger than $1$; $\bar
\kappa\!>\!1$.  However,  $\bar \kappa$ is not included in $S_3$,
which is shown in Fig. \ref{fig-5}(b) ($\bar \kappa$ is included only
in $S_2(=\!\bar \kappa S_3$)) in contrast to the case of $\bar
\gamma$, which is included in $S_1$.  For this reason, $S_3/N_B$ of
the FG model is expected to be smaller than that of the canonical
model. However, the results are opposite; $S_3/N_B$ of the FG model is
slightly larger than that of the canonical model. This result can be
understood from the effective Hamiltonian $\tilde S_{\rm
  eff}\!=\!S_{-1}\!+\! \kappa(\bar\kappa/\bar \gamma) S_3$ described
above. Indeed,  the fact that $\bar\kappa/\bar \gamma <\!1$ makes
$S_3$ larger. In other words, the effective bending rigidity of the FG
model for $\lambda\!=\!0$ corresponds to the slightly smaller $\kappa$
of the canonical model for the same value of $S_3/N_B$. This result is
consistent with the aforementioned result that $\tau$ of the canonical
model for $\kappa\!=\!1$ is linear, whereas $\tau$ of the FG model for
$\kappa\!=\!1$ has a plateau. 

\subsection{FG model for $\lambda\not=0$} 
\begin{figure}[ht]
\centering
\includegraphics[width=9.5cm]{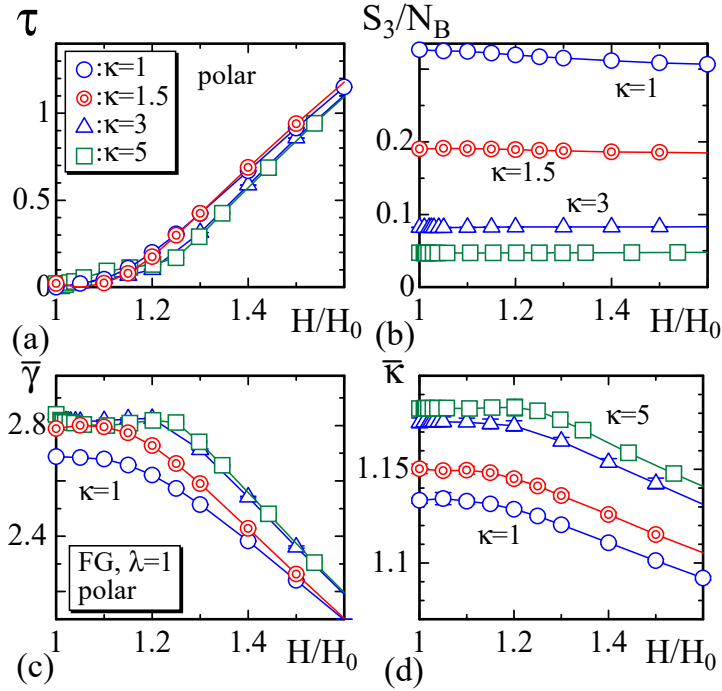}  
\caption{(Color Online) MC data of the polar FG model for
  $\lambda\!=\!1$: (a) $\tau$ vs. $H/H_0$, (b) $S_3/N_B$ vs. $H/H_0$, (c) $\bar \gamma$  vs. $H/H_0$, and (d) $\bar \kappa$  vs. $H/H_0$. $N\!=\!10584$. } 
\label{fig-6}
\end{figure}
Now we turn to the non-trivial cases corresponding to
$\lambda\!\not=\!0$, and we will show that the results, obtained in
the entire range of $\kappa$ including the large $\kappa$ region, are
consistent with the existing J-shaped diagram. First, in
Fig. \ref{fig-6}, we present the results of the polar FG model, where
$\lambda$ is fixed as  $\lambda\!=\!1$. The stress $\tau$ vs. $H/H_0$
in Fig. \ref{fig-6}(a) is found to be J-shaped for $\kappa\!=\!1$,
$\kappa\!=\!1.5$, $\kappa\!=\!3$ and $\kappa\!=\!5$. The result for
$\kappa\!=\!1.5$ in Fig. \ref{fig-6}(a) is new and
non-trivial. Indeed, the corresponding $S_3/N_B$ has values such that
$S_3/N_B\!\simeq\!0.2$, which corresponds to those for
$1.0\!\leq\!\kappa\!\leq\!1.5$ of the canonical model in
Figs. \ref{fig-4}(a), (b), where $\tau$ behaves linearly against
$H/H_0$.  Note that $\tau$ for $\kappa\!\not=\!1.5$ is not always
specific to the FG model because the corresponding $\tau$ also has a
plateau structure just like in the canonical model. The parameters
$\bar \gamma$ and $\bar \kappa$ also have a plateau in the region of $H/H_0$, where $\tau$ has the plateau.   

The problem is why is there no linear behavior of $\tau$ observed in
the FG model. One possible answer is that the effective
one-dimensional correlation introduced by the variable $\sigma$
changes the property of two-dimensional surface fluctuations such that
the long-range force is suppressed in the region of  $H/H_0$ close to
$H/H_0\!=\!1$.  The variable $\sigma$  aligns along the $z$ direction
in which the cylindrical surface is  expanded, and therefore, the
one-dimensional correlation along this direction is expected for a relatively large region of $\lambda$, such as $\lambda\!\geq\!1$. 
Indeed, it is easy to understand from $S_1$ in
Eq. (\ref{2D_discrete_S1}) that a bond length becomes large (small) if
$\sigma$ aligns parallel (vertical) to this bond. Therefore, it is
natural that the surface fluctuations expected in the FG model for the
large  $\lambda$ region are different from those expected in the canonical model. 
This phenomenon in which the long-range force is suppressed is quite analogous to the one reported in Ref. \cite{Koibuchi-PRE2008}, where an  $XY$ model energy suppresses the crumpling transition on spherical surfaces, although the interaction between $\sigma$ and the surface of the XY model in \cite{Koibuchi-PRE2008} is different from the one of the FG model in this paper.

\begin{figure}[ht]
\centering
\includegraphics[width=9.5cm]{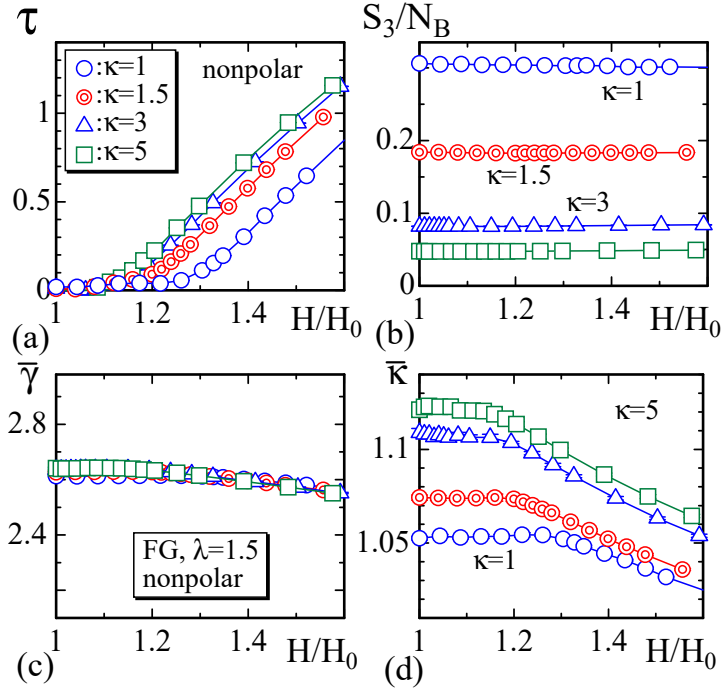}  
\caption{(Color Online) MC data of the nonpolar FG model for
  $\lambda\!=\!1$: (a) $\tau$ vs. $H/H_0$, (b) $S_3/N_B$ vs. $H/H_0$, (c) $\bar \gamma$  vs. $H/H_0$, and (d) $\bar \kappa$  vs. $H/H_0$. $N\!=\!10584$.} 
\label{fig-7}
\end{figure}
The results of the nonpolar FG model are presented in
Fig. \ref{fig-7}, where $\lambda$ is fixed to
$\lambda\!=\!1.5$. These data are consistent with those of the polar
model in the region of $\kappa$ such as $1\!\leq\kappa\!\leq\!5$. For
all values of $\kappa$ assumed, $\tau$ has the J-shaped structure.  

\begin{figure}[ht]
\centering
\includegraphics[width=9.5cm]{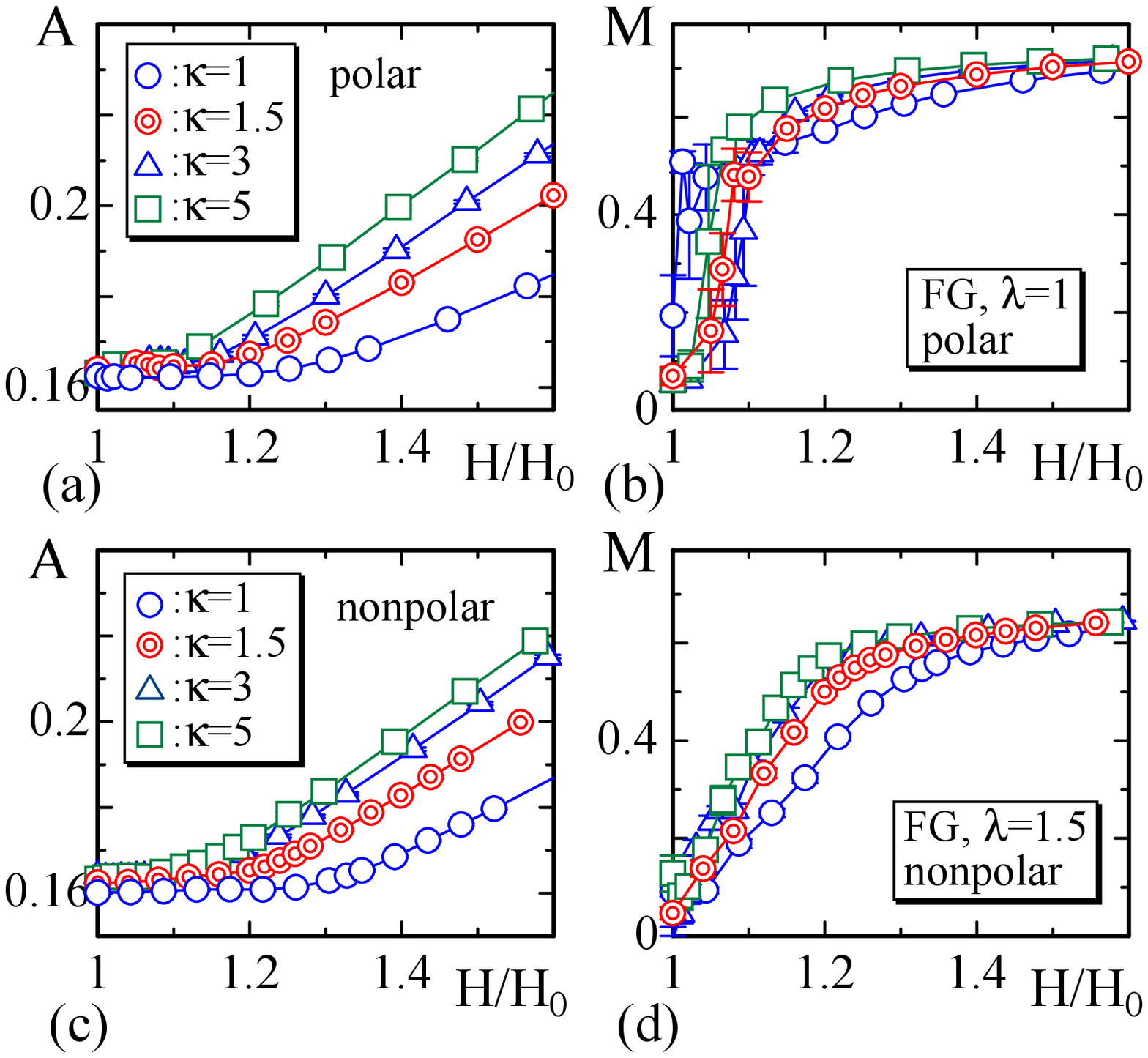}  
\caption{(Color Online) (a) Tangential component $v_{ij}$ of $\sigma_i$ along the direction ${\bf t}_{ij}$ on the triangle $123$, (b) a unit normal vector ${\bf N}_i$ of the tangential plane at the vertex $i$, and the tangential component $\sigma_i^{\mid\mid}$ of $\sigma_i$. } 
\label{fig-8}
\end{figure}
The mean triangle area $A$ and the order parameter $M$ for the variable $\sigma$ defined by 
\begin{eqnarray}
\label{order_param}
M= \left\{ \begin{array}{@{\,}ll}
                  \sigma & \qquad ({\rm polar}) \\
                  (3/2)\left(\sigma_{z}^2-(1/3)\right)  & \qquad ({\rm nonpolar}) 
                  \end{array} 
                   \right.,  
\end{eqnarray}
are plotted in Figs. \ref{fig-8}(a)-(d). In Eq. (\ref{order_param}),
$\sigma $ for the polar case is given by $\sigma\!=\!|\sum_i
\sigma_i|/N$. A plateau can also be detected in $A$, like that in
$\tau$, in both the polar and nonpolar models, and the range $H/H_0$
of the plateau for $A$ is almost identical to that for $\tau$. The
area $A$ corresponds to the real surface area; hence, it is
considerably different from the projected area $A_{\rm p}$, which is
proportional to $H/H_0$. This difference between $A$ and $A_{\rm p}$
implies that the radius of the cylinder shrinks in its plateau region. In fact, it is easy to understand that $A$ has no plateau if the cylinder radius remains unreduced. 

The order parameter changes such that $M\!\to\!0$ ($M\!\to\!1$) for
$H/H_0\!\to\!1$ ($H/H_0\!\to\!\infty$) in
Figs. \ref{fig-8}(b),(d). This result indicates that the origin of the
J-shaped curve is the structural change of $\sigma$. Indeed, $M$
varies rapidly in the plateau region in both the polar and nonpolar models. The plateau of $\tau$ is observed in the range $1\!\leq\!H/H_0\!\leq\!1.2$ ($1\!\leq\!H/H_0\!\leq\!1.3$) for the polar (nonpolar) model. 

\begin{figure}[ht]
\centering
\includegraphics[width=9.5cm]{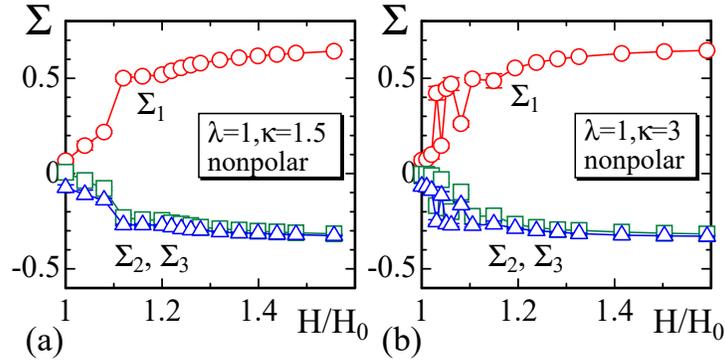}  
\caption{(Color Online) Einegn values $\Sigma_i(i\!=\!1,2,3)$ of the tensor order parameter $Q_{\mu\nu}$ in Eq. (\ref{tensor-order}) of the non-polar model for (a) $\kappa\!=\!1.5$ and (b) $\kappa\!=\!3$, under $\lambda\!=\!1$.} 
\label{fig-9}
\end{figure}
The eigenvalues $\Sigma$ of the tensor order parameter defined by 
\begin{eqnarray}
\label{tensor-order} 
Q_{\mu\nu}=3\left(\langle\sigma_\mu\sigma_\nu\rangle-\delta_{\mu\nu}/3\right)
\end{eqnarray}
are plotted in Fig. \ref{fig-9} for the non-polar model. 
The largest eigenvalue $\Sigma_1$ becomes $\Sigma_1\!\to\! 1$, and the other two eigenvalues $\Sigma_2$ and $\Sigma_3$ are expected to be $\Sigma_{2,3}\!\to\! -0.5$ if $\sigma$ is completely ordered.  We confirm also from Figs. \ref{fig-9} (a),(b) that the variable $\sigma$ becomes ordered if $H$ is enlarged. The behavior of ordering of $\sigma$ is exactly consistent to that of $M$ in Fig. \ref{fig-8} (d). The large fluctuations of $\Sigma_i$ in the small region of $H/H_0$ for $\kappa\!=\!3$ indicate that the directional change of $\sigma$ is abrupt with respect to $H$.

Finally in this subsection, we comment on the reasons for why the bending rigidities of
$\kappa\!\to\!0$ and  $\kappa\!\to\!\infty$ are not assumed in the
calculations. First, the curves of $\tau$ vs. $H/H_0$  are obtained
under the assumption that the surface remains cylindrical in
shape. However, the surface shape deviates from cylindrical and
becomes very thin  for the small $\kappa$ region, such as
$\kappa\!=\!0.5$, and collapses into string-like configurations if
$\kappa$ is reduced to $\kappa\!=\!0.4$. In such a very thin surface,
the surface area becomes far different from the projected
area. Moreover, for these highly wrinkled surfaces, $\tau$ appears to
always be positive even for small $H_0$. This is actually expected because  the surface shrinks to a small sphere for sufficiently small $\kappa$.   
For these reasons, we assume a relatively large bending rigidity ($\kappa\!\geq 0.75$) such that the cylindrical surface shape is maintained in the range $1\!\leq\!H/H_0\!\leq\!2$. In contrast, for the region of large $\kappa$, which is denoted by $\kappa\!\to\!\infty$, $S_2$ is expected to be zero, and therefore, the surface shape can be changed only in its tangential direction. In this case, only $S_{-1}$ changes as $H/H_0$ increases, and there is no reason for $\tau$ to behave non-linearly with respect to $H/H_0$. However, in the case that $S_2$ is not always exactly zero, where the surface is expected to undergo buckling, we also expect a non-linear behavior in $\tau$. However, in these large regions of $\kappa$, the model surface will be far from biological membranes.

\subsection{Comparison  with experimental data} \label{comparison}
\begin{figure}[ht]
\centering
\includegraphics[width=9.5cm]{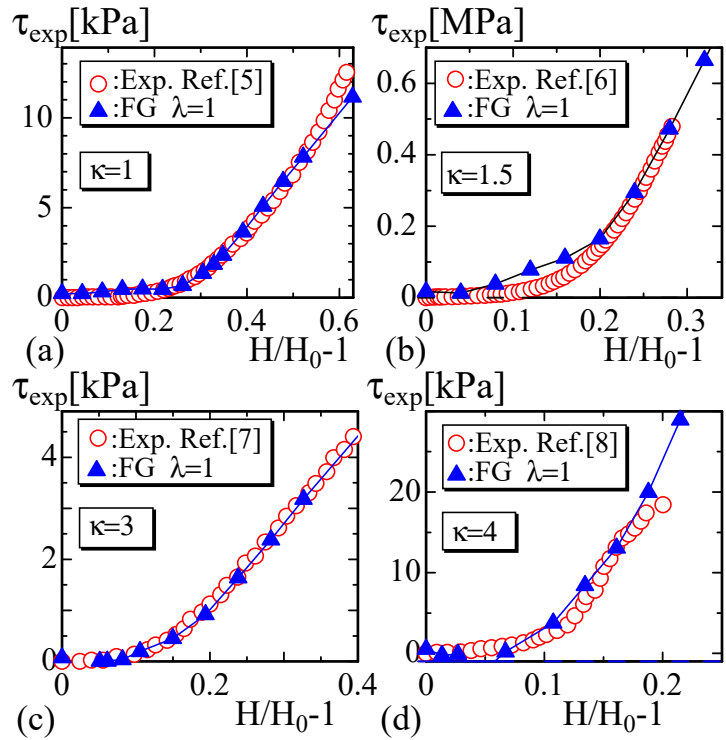}  
\caption{(Color Online) The nominal stress vs. strain (${\color{red} \bigcirc}$) of (a)  blood vessel \cite{TDRW-JMCB2013}, (b) rat muscle \cite{TFCC-CLINICS2010}, (c) collagen fibers \cite{RKSRV-ASME2002} and (d) collagen hydrogels \cite{SBVN-ABME2000}. The solid triangle denotes the simulation data $\tau_{\rm sim}$ of Eq. (\ref{experimetal_tension}) for the non-polar interaction.} 
\label{fig-10}
\end{figure}
In this subsection, we show that the simulation data can be compared to experimental stress-strain data of biological materials such as blood vessel, rat muscle, collagen fibers, and collagen hydrogels \cite{TDRW-JMCB2013,TFCC-CLINICS2010,RKSRV-ASME2002,SBVN-ABME2000} (see Figs. \ref{fig-10}(a)--(d)). 
First, we should comment on the unit of $\tau$ in Eq. (\ref{frame_tension}) assumed for the simulations in more detail. In the simulations, the inverse temperature $\beta(=\!1/k_BT)$  is fixed to $\beta\!=\!1(\Leftrightarrow k_BT\!=\!1)$, and under this unit the triangle edge length $a$ is fixed to $a\!=\!1$. This $a$ corresponds to the lattice spacing in the lattice field theory language \cite{Creutz-txt} and is suitably fixed such that the simulation data can be compared to the experimental data. However, the physical unit of $\tau$ is given by $[N/m]$, which is different from the experimental one $[Pa]$ for stresses of macroscopic objects. For this reason, we obtain $\tau_{\rm sim}$ dividing the simulation data $\tau$ by $a$ to compare $\tau$ with the experimental stresses.  By including $\beta$, $a$ and $k_BT$ in the calculation formula of $\tau$, we have
\begin{eqnarray} 
\label{experimetal_tension}
\tau_{\rm sim} &=&  \frac{2\gamma \langle S_1\rangle - 3N+2N_2}{2A_{\rm p}}\frac{k_BT}{a^3} \nonumber \\
&=&\left({4\times 10^{-21}}/{a^3}\right)\tau, 
\end{eqnarray}
where the room temperature is assumed for $T$.   Note that the simulation data $\tau$ in Eq. (\ref{frame_tension}) is obtained from this $\tau_{\rm sim}$ by assuming $k_BT\!=\!1$ and $a\!=\!1$. Note also that the unit of $\left({4\times 10^{-21}}/{a^3}\right)$ is $[m^{-1}]$ because the units of $\tau_{\rm sim}$ and $\tau$ are given by $[N/m^2]$ and $[N/m]$, respectively. This $\tau_{\rm sim} [N/m^2]$ can be identified with experimental stresses if the value of $a$ is specified. The problem is how to obtain the coefficient $\left({4\times 10^{-21}}/{a^3}\right)$ from experimental and simulation data. One possible answer is to determine $\left({4\times 10^{-21}}/{a^3}\right)$ such that the slope of $\tau_{\rm sim}$ equals to that of experimental data in their linear regions. The slope of $\tau$ with respect to the strain is just Young modulus $E_{\rm sim}$. Therefore, the experimental and simulation  Young moduluses $E_{\rm exp}$ and  $E_{\rm sim}$ can be obtained from their linear part of  the corresponding experimental nominal stress and $\tau$ such that the following condition is satisfied:
\begin{eqnarray} 
\label{experimetal_Young_modulus}
E_{\rm exp} &=&  \left({4\times 10^{-21}}/{a^3}\right) E_{\rm sim}.
\end{eqnarray}
From this relation, the coefficient $\left({4\times 10^{-21}}/{a^3}\right)$ is obtained and used to plot $\tau_{\rm sim}$ in Figs. \ref{fig-10}(a)-(d).  We find in the experimental data that the linear behavior terminates for large strain region (see  Fig. \ref{fig-10}(d)). In contrast, the simulation data $\tau_{\rm sim}$ behave only linearly for large strain region because no failure mechanism is implemented in the model.

\begin{table}[htb]
\caption{The lattice spacing $a$ corresponding to the stress-strain diagrams shown in Figs. \ref{fig-10}(a)--(d). }
\label{table-1}
\begin{center}
 \begin{tabular}{ccccccccc}
 \hline
 Fig.\ref{fig-10} && (a) && (b) && (c) && (d)    \\
 \hline
 $a[m]$ &&  $6.9\times 10^{-9}$ && $1.30\times 10^{-8}$  && $8.53\times 10^{-9}$  && $3.14\times 10^{-9}$ \\
 \hline
 \end{tabular} 
\end{center}
\end{table}
From the coefficients $\left({4\times 10^{-21}}/{a^3}\right)$, which are obtained from the experimental and simulation data and used for the plots in Figs. \ref{fig-10}(a)--(d), the parameters $a$ can be obtained and are shown in  Table \ref{table-1}. The parameters $a$ are approximately $10$ times (or more) greater than the Van der Waals radius of atoms, and therefore these $a$ are meaningful as the lattice spacing for the calculations of $\tau_{\rm sim}$. 

\section{Summary and conclusion}
We have studied the origin of the J-shaped stress-strain diagram using
Monte Carlo simulations on triangulated surfaces. For such a
non-linear behavior of the J-shaped diagram, it has been widely
accepted that the collagen structure plays an essential role
\cite{Maier-Saupe-1958,de-Gennes-1979,Doi-Edwards-1986,JMBBM-2016,MCPolymer-PRE2008,Polymer-PRE2003,Fiber-RMP2010,Fiber-PRE2012}. However,
for the J-shaped diagram, no concrete result has yet been obtained in
theoretical or computational evaluations of the curve from the
perspective of two-dimensional surface models because the interaction
between the collagen and the bulk material (including collagen itself) is too complex. 

To understand the mechanism of the J-shaped diagram, we first
calculate  the frame tension $\tau$ of cylindrical surfaces  using the
canonical surface model of Helfrich and Polyakov (HP)
\cite{BOWICK-TRAVESSET-EPJE2001,Cuerno-etal-2016,Koibuchi-etal-JOMC2016}.
From the Monte Carlo data of the HP model, we find that $\tau$ is
J-shaped. However, the J-shaped curve can be obtained only for some
limiting cases, such as small and large bending rigidity $\kappa$
regions. In fact, for the region of $\kappa\!\simeq\! 1.5$, $\tau$
changes linearly with respect to $H/H_0$, including the smaller region
$H/H_0\!\simeq\!1$. For this reason, we apply the Finsler geometry
(FG) model to evaluate $\tau$ on the same cylindrical surfaces. The FG
model is an extension of the HP model and includes a new degree of
freedom $\sigma$ corresponding to the polymer (or liquid crystal)
direction
\cite{Koibuchi-Sekino-PhysicaA2014,OK-POL2017,KS-IJMPC2016}. The Monte
Carlo results of the FG model for all values of $\kappa$ are in good
agreement with the existing J-shaped stress-strain curves obtained experimentally. This result implies that the J-shaped diagram can be understood in the context of the FG modeling. 

The important point to note is that a structural change is essential
for the J-shaped curve. This structural change is associated with the
directional degrees of freedom of $\sigma$, which has two different
phases, such as ordered and disordered. A phase transition between
these two phases, from the disordered phase to the ordered phase, is
activated by an external force that expands the surface. In this
expansion process, the external force changes the internal structure
represented by $\sigma$ in the small $H/H_0$ region. As a result of
this structural change,  the surface fluctuation property is altered
such that a long-range correlation, expected for a certain region of $\kappa$ in the canonical model, is suppressed due to the one-dimensional correlation of $\sigma$. 
Thus, the internal structural change during the process of surface
expansion is the origin of the J-shaped stress-strain diagram of
membranes, and this intuitive picture for the interaction between
$\sigma$ and the bulk polymer can be implemented in the FG surface
model. We should note that the detailed information on the transition
property of this internal structure and the dependence  of the J-shaped curve on the internal phase transition remain to be studied.

\acknowledgements
The author H.K. acknowledges Giancarlo Jug and Andrei Maximov for discussions and comments. The authors acknowledge Eisuke Toyoda for computer analyses. This work is supported in part by JSPS KAKENNHI Numbers 26390138 and 17K05149.

\appendix

\section{Finsler geometry model for 2D membrane}\label{FGmodel_Appendix}
We start with the continuous Hamiltonian $S$, which is given by
\begin{eqnarray}
\label{cont_S}
&& S=\gamma S_1+\kappa S_2, \nonumber \\
&&S_1=\int \sqrt{g}d^2x g^{ab} \frac{\partial {\bf r}}{\partial x_a}\cdot \frac{\partial {\bf r}}{\partial x_a}, \nonumber \\
&&S_2=\frac{1}{2}\int \sqrt{g}d^2x  g^{ab} \frac{\partial {\bf n}}{\partial x_a} \cdot \frac{\partial {\bf n}}{\partial x_b},  
\end{eqnarray} 
where $S_1$ and $S_2$ are the Gaussian energy and the bending energy,
respectively. The coefficients $\gamma(=\!1)$ and  $\kappa[1/k_BT]$
are the surface tension and the bending rigidity, respectively, where
$k_B$ and $T$ are the Boltzmann constant and the temperature. In
$S_1$, ${\bf r} (\in {\bf R}^3)$ is the surface position (see
Fig. \ref{fig-2}(b)), which is locally parametrized by $(x_1,x_2)$;
hence, ${\bf r}$ is understood to be a mapping from the
two-dimensional parameter space $M$ to ${\bf R}^3$ such that $M \ni
(x_1,x_2)\mapsto {\bf r}(x_1,x_2) \in {\bf R}^3$.  The $2\!\times\! 2$
matrix $g_{ab}$ is a metric function on $M$, $g^{ab}$ is its inverse,
and $g$ is the determinant of $g_{ab}$. The symbol ${\bf n}$ in $S_2$
is a unit normal vector of the surface ${\bf r}(x_1,x_2)$ in ${\bf
  R}^3$ (see Ref. \cite{Koibuchi-Sekino-PhysicaA2014} for more
details).  The FG is considered to be a framework for anisotropic phenomena \cite{Matsumoto-SKB1975,Bao-Chern-Shen-GTM200,George-Bogoslovsky-JGM2012,George-Bogoslovsky-PLA1998}.   
We also note that the J-shaped diagram is expected to share the same origin with  the soft elasticity in 3D liquid crystal elastomers \cite{Wamer-Terentjev,V-Domenici-2012,Terentjev-JPCM-1999,K-F-MacMolCP-1998}. 

Let $\sigma_i (\in S^2: {\rm unit\;sphere})$ be the variable
corresponding to the directional degrees of freedom  of a polymer or
molecule such as liquid crystals at the vertex $i$ of the triangulated surface. Let ${\bf t}_{ij}$ be the unit tangential vector of the triangle edge (or bond) $ij$, which connects the vertices $i$ and $j$, such that 
${\bf t}_{ij}=({\bf r}_j-{\bf r}_i)/|{\bf r}_j-{\bf r}_i|$. 
Using this ${\bf t}_{ij}$, we define the tangential component of $\sigma_i$ along the bond $ij$ by
\begin{eqnarray}
\label{tangent_comp}
v_{ij}=\left|\sigma_i \cdot {\bf t}_{ij}\right|.
\end{eqnarray} 
We note that $v_{ij}\not=v_{ji}$ in general.

Let $123$ denote a triangle on $M$, and let the vertex $1$ be the
local coordinate origin of the triangle $123$; then, the Finsler metric $g_{ab}$ on the triangle $123$ is defined by 
\begin{equation}
\label{Finsler_metric}
g_{ab}=\left(  
       \begin{array}{@{\,}cc}
         1/v_{12}^{2}  & 0\\
         0  & 1/v_{13}^{2}  
        \end{array} 
       \\ 
 \right).
\end{equation}
Thus, by the replacements
\begin{eqnarray}
\label{discretization}
&&\partial_1  {\bf r} \to {\bf r}_2- {\bf r}_1, \quad
\partial_2  {\bf r} \to {\bf r}_3- {\bf r}_1,   \nonumber \\
&&\partial_1  {\bf n} \to {\bf n}_0- {\bf n}_2, \quad
\partial_2  {\bf n} \to {\bf n}_0- {\bf n}_3, \nonumber \\
&&\int \sqrt{g}d^2x \to \frac{1}{2}\sum_{\Delta} \sqrt{\det g_{ab}}
\end{eqnarray}
in Eq. (\ref{cont_S}), we have 
\begin{eqnarray} 
\label{2D_discrete_S1}
&&S_1=\frac{1}{2}\sum_{\it \Delta} \left[ \frac{v_{12}}{v_{13}} \ell_{12}^2 + \frac{v_{13}}{v_{12}} \ell_{13}^2\right],\; \ell_{ij}^2=({\bf r}_i- {\bf r}_j)^2, \\
&&S_2=\frac{1}{2}\sum_{\it \Delta} \left[ \frac{v_{13}}{v_{12}} (1-{\bf n}_0\cdot {\bf n}_3) + \frac{v_{12}}{v_{13}} (1-{\bf n}_0\cdot {\bf n}_2)\right].\nonumber 
\end{eqnarray}

\begin{figure}[ht]
\centering
\includegraphics[width=9.5cm]{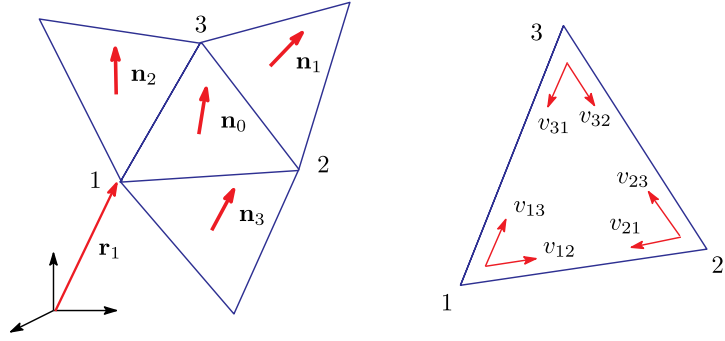}  
\caption{(Color Online) (a) Tangential component $v_{ij}$ of $\sigma_i$ along the direction ${\bf t}_{ij}$ on the triangle $123$, (b) the triangle $123$ and the three neighboring triangles with unit normal vectors ${\bf n}_i$, $(i=0,1,2,3)$, ${\bf r}_1$ is the position of vertex $1$.
 } 
\label{fig-A1}
\end{figure}
Because there are three possible local coordinate origins on the triangle $123$, all possible terms in $S_1$ and $S_2$ should be summed over with the coefficient $1/3$.  The sum over triangles $\sum_{\it \Delta}$ in both $S_1$ and $S_2$ can be replaced by the sum over bonds $\sum_{ij}$; then, we finally obtain   
\begin{eqnarray} 
\label{2D_discrete_S1_triangle}
&&S_1=\frac{1}{6}\sum_{\it \Delta} \left[ \gamma_{12} \ell_{12}^2 + \gamma_{23} \ell_{23}^2  + \gamma_{31} \ell_{31}^2\right], \nonumber \\
&&S_2=\frac{1}{6}\sum_{\it \Delta} \left[ \kappa_{12} \left(1-{\bf n}_0\cdot {\bf n}_3\right) + \kappa_{23} \left(1-{\bf n}_0\cdot {\bf n}_1\right) \right. \nonumber \\
&& \qquad\qquad\qquad + \left. \kappa_{31} \left(1-{\bf n}_0\cdot {\bf n}_2\right)\right],  \\
&&\gamma_{12}=\frac{v_{12}}{v_{13}} + \frac{v_{21}}{v_{23}}, \; \gamma_{23}=\frac{v_{23}}{v_{21}} + \frac{v_{32}}{v_{31}},  \; \gamma_{31}=\frac{v_{31}}{v_{32}}+ \frac{v_{13}}{v_{12}}, \nonumber \\
&&\kappa_{12}=\frac{v_{13}}{v_{12}} + \frac{v_{23}}{v_{ 21}},  \; \kappa_{23}=\frac{v_{21}}{v_{23}} + \frac{v_{31}}{v_{32}}, \; \kappa_{31}= \frac{v_{32}}{v_{31}} + \frac{v_{12}}{v_{13}}. \nonumber
\end{eqnarray}
Multiplying  $\gamma (=\!1)$ by $\gamma_{ij}$ and $\kappa$ by
$\kappa_{ij}$, we have $\gamma\gamma_{ij}$ and $\kappa\kappa_{ij}$,
which can be considered to be effective surface tension and effective bending rigidity. These quantities $\gamma\gamma_{ij}$ and $\kappa\kappa_{ij}$ are dependent on the position and the direction of the bond $ij$, although $\gamma_{ij}$ and $\kappa_{ij}$ are a part of energies $S_1$ and $S_2$, respectively. This dependence of $\gamma_{ij}$ and $\kappa_{ij}$ on the position and the direction of the bond is the most interesting output of the FG model. Anisotropic coefficients are expected to play an important role for the anisotropy in LCE \cite{Lubensky-PRE-2002,Lubensky-PRE-2003,Xing-Radzihovsky-ANP-2008,Stenul-Lubensky-PRL-2005}. Note that both $S_1$ and $S_2$ are explicitly dependent on $\sigma$ because $\gamma_{ij}$ and $\kappa_{ij}$ are determined by $\sigma$ via Eq. (\ref{tangent_comp}).


\begin{thebibliography}{0}

\bibitem{Biomat-2013}
J. P. Chowa, D. T. Simionescu, H. Warner, B. Wang, S. S. Patnaik, J. Liao, and A, Simionescu, Biomaterials {\bf 34},  pp.685-695 (2013).

\bibitem{MCLS-PMatSAc-2008}
M.A. Meyers, P. Chen, A.Y. Lin, and Y. Seki, Prog. Mat. Science, {\bf 53}, pp.1-206 (2008). 

\bibitem{Greven-etal-JMorph-1995}
H. Greven, K. Zanger, and G. Schwinger, J. Morphology, {\bf 224}, pp.15-22 (1995). 

\bibitem{FMZRAB-JSBiol-1997}
P. Fratzl, K. Misof, I. Zizak, G. Rapp, H. Amenitsch, and S. Bernstorff, J. Str. Biol. {\bf 122}, pp.119-122, SB983966 (1997).

\bibitem{TDRW-JMCB2013}
 G. Tronci, A. Doyle, S. J. Russell and  D.J. Wood,  J. Mater. Chem. B {\bf 1}, 5478 (2013).

\bibitem{TFCC-CLINICS2010}
A.E. Toscano, K.M. Ferraz, R.M. de Castro and F. Canon, Clinics {\bf 65}, 1363 (2010).

\bibitem{RKSRV-ASME2002}
K. Kokini, J.E. Sturgis, J.P. Robinson and S.L.Voytik-Harbin, J. Biom. Eng., Trans. ASME, {\bf 124}, 214 (2002).

\bibitem{SBVN-ABME2000}
D. Seliktar, R.A. Black, R.P. Vito, and R.M. Nerem, Annals Biom. Eng. {\bf 28}, 351 (2000).



\bibitem{Flory-1959}
P. J. Flory, \textit{Principles of Polymer Chemistry}, (Cornell University, Ithaca, 1953).

\bibitem{JMBBM-2013}
B. Xua, Y. Lia, X. Fanga, G. A. Thouasb, W. D. Cooka, D. F. Newgreenc, and Q. Chena, 
J. Mech. Behav. Biomed. Mat. {\bf 28}, pp.354-365 (2013).

\bibitem{GVRB-NL2011}
A. Gautieri, S. Vesentini, A. Redaelli, and M.J. Buehler, Nano Lett. {\bf 11}, 757 (2011).



\bibitem{Maier-Saupe-1958}
W. Maier and A. Saupe, Z. Naturforsch, {\bf 13A}, 564 (1958); {\bf 14A}, 882 (1959);  {\bf 15A}, 287 (1960).

\bibitem{de-Gennes-1979}
P.-G.de Gennes, \textit {Scaling Concepts in Polymer Physics}, (Ithaca-London, Cornell Univ. Press, 1979).

\bibitem{Doi-Edwards-1986}
M. Doi and S.F. Edwards, \textit {The Theory of Polymer Dynamics}, (Oxford University Press, 1986).

\bibitem{JMBBM-2016}
A. Levillain, M. Orhant, F. Turquier, and T. Hoc, 
J. Mech. Behav. Biomed. Mat. {\bf 61}, pp.308-317 (2016).

\bibitem{MCPolymer-PRE2008}
E. M. Huisman, C. Storm, and G. T. Barkema, Phys. Rev. E {\bf 78}, 051801 (2008).

\bibitem{Polymer-PRE2003}
D. A. Head, A. J. Levine, and F. C. MacKintosh, Phys. Rev. E {\bf 68}, 061907 (2003).

\bibitem{GJug-JNCS2014}
G. Jug, M. Paliienko, S. Bonfanti, J. Non-Cryst. Solids {\bf 401}, pp.66-72 (2014).

\bibitem{Fiber-RMP2010}
S. Pradhan, A. Hansen, B. K. Chakrabarti,  Rev. Mod. Phys. {\bf 86}, pp.499-555 (2010).

\bibitem{Fiber-PRE2012}
T. Giesa, N. M. Pugno, and M. J. Buehler, Rev. Mod. Phys. {\bf 82}, 041902 (2012).

\bibitem{HELFRICH-1973}
 W. Helfrich, Z. Naturforsch {\bf 28c}, 693 (1973).

\bibitem{POLYAKOV-NPB1986}
 A.M. Polyakov, Nucl. Phys. B {\bf 268}, 406 (1986).

\bibitem{Bowick-PREP2001}
 M. Bowick and A. Travesset, Phys. Rep. {\bf 344}, 255 (2001).

\bibitem{WIESE-PTCP19-2000}
 K.J. Wiese, \textit{Phase Transitions and Critical Phenomena 19}, 
 edited by C. Domb, and  J.L. Lebowitz (Academic Press, 2000) p.253.

\bibitem{NELSON-SMMS2004}
D. Nelson, in \textit {Statistical Mechanics of Membranes and Surfaces}, Second Edition, edited by  D. Nelson, T. Piran, and S. Weinberg, (World Scientific, 2004) p.1. 

\bibitem{GOMPPER-KROLL-SMMS2004}
G. Gompper and D.M. Kroll, Triangulated-surface models of fluctuating membranes, in {\it Statistical Mechanics of Membranes and Surfaces, Second Edition}, eds.  D. Nelson, T. Piran, and S. Weinberg, (World Scientific, Singapore, 2004) p.359. 

\bibitem{Koibuchi-Sekino-PhysicaA2014}
 H. Koibuchi and H. Sekino, Physica A {\bf 393}, 37 (2014).

\bibitem{OK-POL2017}
 K. Osari and H. Koibuchi, Polymer {\bf 114}, 355 (2017).

\bibitem{WHEATER-JP1994}
 J.F. Wheater, J. Phys. A Math. Gen. {\bf 27}, 3323 (1994).

\bibitem{Cai-Lub-PNelson-JFrance1994}
W. Cai, T. C. Lubensky, P. Nelson, and T. Powers, J. Phys. II France {\bf 4},  p.931 (1994)

\bibitem{David-Leibler-JPF1991}
F. David and S. Leibler, J. Phys. II Frans {\bf 1}, pp.959-976 (1991) 

\bibitem{Wamer-Terentjev}
M. Wamer, and E.M. Terentjev, \textit {Liquid Crystal Elastomer}, (Oxford University Press, 2007). 

\bibitem{V-Domenici-2012}
V. Domenici, Prog. Nucl. Mag. Res. Spec. {\bf 63}, 1 (2012).

\bibitem{Terentjev-JPCM-1999}
E.M. Terentjev, J. Phys. Condens. Matter {\bf 11},  R239 (1999).

\bibitem{K-F-MacMolCP-1998}
I. Kundler, H. Finkelmann, Macromol. Chem. Phys. {\bf 199}, 677 (1998).

\bibitem{KS-IJMPC2016}
H. Koibuchi and A. Shobukhov, {\bf 27}, 1650042(1-15) (2016); Erratum {\bf 27}, 1692001(1-1) (2016).

\bibitem{Leb-Lash-PRA1972}
 P. A. Lebwohl and G. Lasher, Phys. Rev. A {\bf 6}, 426 (1972).

\bibitem{Koibuchi-etal-JOMC2016}
H. Koibuchi, A. Shobukhov and H. Sekino, J. Math. Chem. {\bf 54}, 358 (2016). 

\bibitem{Creutz-txt}
M. Creutz, \textit {Quarks, gluons and lattices}, (Cambridge University Press, Cambridge, 1983. 

\bibitem{Mepropolis-JCP-1953}
N. Metropolis,  A. W. Rosenbluth, M. N. Rosenbluth, A. H. Teller, J. Chem. Phys. {\bf 21}, pp.1087-1092 (1953).

\bibitem{Landau-PRB1976}
D.P. Landau, Phys. Rev. B {\bf 13}, pp.2997-3011 (1976).

\bibitem{Koibuchi-PRE2008}
H. Koibuchi, Phys. Rev. E. {\bf 77}, 021104(1-8) (2008).

\bibitem{Matsumoto-SKB1975}
M. Matsumoto, {\it Keiryou Bibun Kikagaku} (in Japanese), (Shokabo, Tokyo, 1975).

\bibitem{Bao-Chern-Shen-GTM200}
D. Bao, S. -S. Chern, Z. Shen, {\it An Introduction to Riemann-Finsler Geometry}, (Springer, New York, 2000).

\bibitem{George-Bogoslovsky-JGM2012}
G. Bogoslovsky, Int. J. Geom. Methods Mod. Phys. {\bf 9},  1250007 (2012).

\bibitem{George-Bogoslovsky-PLA1998}
G. Bogoslovsky, Phys. Lett. A {\bf 244}, 222 (1998).



\bibitem{BOWICK-TRAVESSET-EPJE2001}
M. J. Bowick, A. Cacciuto, G. Thorleifsson, and A. Travesset, Eur. Phys. J. E {\bf 5}, 149 (2001).


\bibitem{Cuerno-etal-2016}
R. Cuerno, R. Gallardo Caballero, A. Gordillo-Guerrero,  P. Monroy, and J. J. Ruiz-Lorenzo, Phys. Rev. E {\bf 93}, 022111(1-9) (2016).

\bibitem{Lubensky-PRE-2002}
T. C. Lubensky, R. Mukhopadhyay, L. Radzihovsky and X. Xing, Phys. Rev. E {\bf 66}, 011702 (2002).

\bibitem{Lubensky-PRE-2003}
X. Xing, R. Mukhopadhyay, T. C. Lubensky, and L. Radzihovsky, Phys. Rev. E {\bf 68}, 021106 (2003).

\bibitem{Xing-Radzihovsky-ANP-2008}
X. Xing and L. Radzihovsky, Annals of Phys. {\bf 323}, 105 (2008).

\bibitem{Stenul-Lubensky-PRL-2005}
O. Stenull and T. C. Lubensky, Phys. Rev. Lett. {\bf 94}, 018304 (2005).

\end{thebibliography}
\end{document}